\documentclass[twocolumn, twocolappendix]{aastex631}
\usepackage{mathrsfs}
\usepackage{amsmath}
\numberwithin{equation}{section}
\usepackage{graphicx}
\usepackage{times}
\usepackage[varg]{txfonts}
\usepackage{hyperref}
\usepackage{multirow}
\usepackage{xcolor}
\usepackage{array}

\newcommand{\blu}{\mathcal{B}^\mathrm{L}_\mathrm{U}}
\newcommand{\HL}{\mathcal{H}_\mathrm{L}}
\newcommand{\HU}{\mathcal{H}_\mathrm{U}}

\newcommand{\olu}{\mathcal{O}^\mathrm{L}_\mathrm{U}}
\newcommand{\plu}{\mathcal{P}^\mathrm{L}_\mathrm{U}}

\newcommand{\fdp}{\mathrm{FDP}}
\newcommand{\fpp}{\mathrm{FPP}}
\newcommand{\plensed}{p_\mathbb{L}}
\newcommand{\purity}{c_\mathbb{L}}

\newcommand{\pdet}{P_\mathrm{det}}

\newcommand{\theq}{\theta_\mathrm{eq}}
\newcommand{\thbI}{\theta_\mathrm{b1}}
\newcommand{\thbII}{\theta_\mathrm{b2}}
\newcommand{\dthb}{\Delta\theta_\mathrm{b}}
\newcommand{\dphi}{\Delta\phi}
\newcommand{\dt}{\Delta t}

\begin{document}

\title{Search for strong lensing of gravitational waves in the binary black hole events from O1--O4a}

\author{Ankur Barsode$^a$}
\email{$^a$ankur.barsode@icts.res.in}
\affiliation{International Centre for Theoretical Sciences, Tata Institute of Fundamental Research, Bangalore 560089, India}

\author{Koustav N. Maity$^{b}$}
\email{$^b$koustav.narayan@icts.res.in}
\affiliation{International Centre for Theoretical Sciences, Tata Institute of Fundamental Research, Bangalore 560089, India}

\author{Parameswaran Ajith$^{c}$}
\email{$^c$ajith@icts.res.in}
\affiliation{International Centre for Theoretical Sciences, Tata Institute of Fundamental Research, Bangalore 560089, India}

\begin{abstract}
A small fraction of the gravitational waves (GWs) currently observable by LIGO, Virgo, and KAGRA (LVK) may be strongly lensed by intervening galaxies and galaxy clusters, potentially producing multiple copies of the same signal. We search for lensed pairs of binary black hole signals detected during the O1-O4a observing runs. We include the events identified by the LVK Collaboration, as well as additional events found by external groups (IAS and OGC). Our search is based on \textsc{Posterior Overlap 2.0}, a fast and efficient Bayesian model-selection pipeline to identify lensed candidates. The search is supplemented by realistic background and foreground simulations to characterize the robustness and detection efficiency of the pipeline, as well as the statistical significance of lensed candidates. We define new metrics to assess the statistical significance of lensing both at the individual and population levels. Our work addresses some of the limitations of previous searches. With the probability of lensing $<0.6\%$ for all pairs, we find no evidence for strong lensing in the data and consequently place a 90\% upper bound on the lensing fraction of 1.4\%. With five out of the top nine lensed candidate pairs being from non-LVK catalogs, we also highlight the importance of searching among events reported by multiple GW catalogs. We forecast that the probabilities of making a $3\sigma$ detection in the fourth (O4), intermediate (IR1), and fifth (O5) observing runs are $\sim 20\%,\, 23\%$, and 67\%, respectively.
\end{abstract}

\section{Introduction}
\label{sec:introduction}
According to general relativity (GR), electromagnetic waves and gravitational waves (GWs) are deflected by massive objects in their propagation path. This phenomenon, gravitational lensing, has emerged as a powerful tool for astrophysics and cosmology from the observations of electromagnetic waves. While gravitational lensing of GWs has not yet been detected, this is expected in the upcoming observation runs of LIGO-Virgo-KAGRA (LVK)~\citep{barsode2026lensing}.

In the case of massive lenses such as galaxies and galaxy clusters, where the gravitational radius is much larger than the wavelength of the radiation, lensing effects can be approximated by geometric optics. In \emph{strong lensing}, multiple copies (i.e., images) of the same signal will be observed, where each copy may be magnified, phase shifted, and time delayed with respect to the original unlensed signal, mostly without altering its frequency evolution.

Once detected, strongly lensed GWs can significantly advance our understanding of galaxies~\citep{seo2024inferring, xu2022please}, the nature of dark matter~\citep{jana2025probing}, and cosmology~\citep{Liao:2017ioi,Wei:2017emo,jana2023cosmography,jana2024strong,chen2026forecasting, chen2026prospect, maity2026strong}. They can also be used to test various predictions of GR~\citep{Goyal:2020bkm, Goyal:2023uvm, Collett:2016dey, Fan:2016swi}. Strongly lensed events may also enable tighter sky localisation of the sources, potentially enabling us to identify their host galaxy~\citep{wempe2024detection, hannuksela2020localizing, ng2025uncovering, uronen2025finding}.

Roughly $0.1\%$ of GWs---mainly binary black holes (BBHs)---observable by the LVK detectors \citep{aasi2015advanced, acernese2014advanced,akutsu2021overview,aso2013interferometer,somiya2012detector} may be strongly lensed by galaxies and galaxy clusters (see Sec.~\ref{sec:priors}, and also, e.g., \cite{ng2018precise, xu2022please, wierda2021beyond, mukherjee2021impact, maity2026strong}). While a $3\sigma$ detection is likely during the fifth observing of LVK~\citep{barsode2026lensing}, the chances of finding a lensed GW in the current data are no longer negligible. With $\sim 200$ GW detections confirmed so far~\citep{abbott2019gwtc, abbott2021gwtc, abbott2023gwtc, abbott2024gwtc, abac2025gwtc, venumadhav2020new, zackay2021detecting, olsen2022new, mehta2025new, wadekar2023new, nitz20191, nitz20202, nitz20213, nitz20234, koloniari2025new}, a non-detection could also be useful for ruling out some of the extreme models of the BBH merger rate~\citep{harshe2026could}, their formation channels~\citep{leong2024constraining}, or the nature of dark matter~\citep{basak2022constraints,barsode2024constraints}.

Since strongly lensed GWs are quite similar to unlensed GWs, a strong lensing search essentially involves identifying pairs of GW signals with similar morphology in their polarisations (apart from an overall magnification and phase shift), coming from the same region in the sky, with time delays and magnification ratios consistent with our astrophysical expectations. As in any signal detection problem, there is a chance of obtaining false alarms when two unrelated (unlensed) signals show some accidental consistency. The optimal method for separating truly lensed signals from such false alarms is the Bayesian model selection, which involves computing the Bayes factor (i.e., the likelihood ratio) between the two competing hypotheses (lensed vs. unlensed) from each pair of GW events. Alternative methods, based on cross-correlation~\citep{chakraborty2024glance, Kopty:2026ugr} and machine learning~\citep{goyal2021rapid, magare2024slick, sun2026identifying, li2026identification, zhang2026time} have also been proposed to detect strong lensing.

Unfortunately, computing the Bayes factor by simultaneously analysing all pairs of GW events is computationally prohibitive~\citep{janquart2021fast, janquart2023return, lo2023bayesian}. Thus, the current LVK searches~\citep{LIGOScientific:2021izm, abbott2023search, janquart2023follow, abac2025gwtclens} have used sub-optimal, but computationally inexpensive detection statistics~\citep{haris2018identifying, goyal2021rapid, ezquiaga2023identifying}\footnote{Some of these statistics can be thought of as various approximations of the full Bayes factor~\citep{barsode2025fast}} to analyse all pairs. Only the event pairs that have the highest ranks in the first round are analyzed by the joint parameter estimation pipelines to compute the full Bayes factor.

In the searches conducted so far, no statistically significant candidates have been found~\citep{hannuksela2019search, dai2020search, liu2021identifying, LIGOScientific:2021izm, abbott2023search, janquart2023follow, abac2025gwtclens}. However, this non-detection has a few caveats: apart from the suboptimality of the first stage of the search, the second stage is too expensive to perform a large number of background simulations to estimate the statistical significance of the top candidates. Secondly, these searches only include event pairs within the same observing run and ignore the pairs that can be formed between observing runs. Finally, they ignore the BBH candidates identified by groups outside LVK~\citep{venumadhav2020new, zackay2021detecting, olsen2022new, mehta2025new, wadekar2023new,nitz20191, nitz20202, nitz20213, nitz20234}; some of these groups have expanded the range of BBH searches to higher masses, where we are more likely to identify lensing~\citep{olsen2022new, mehta2025new,cheung2023mitigating}.

These caveats make it difficult to estimate the uncertainty in the null result reported by these searches. This is important because a majority of truly lensed pairs may be buried within false alarms~\citep{ccalicskan2023lensing, barsode2026lensing}. If unaccounted for, these would bias the results of various downstream studies based on the current non-observation. In addition, in the absence of any highly statistically significant lensing candidates, these uncertainties may hold the key to detecting the presence of lensing in the data at a population level.

In this work, we address these limitations by performing a single-stage strong lensing search between all the BBH events detected so far from the first three observing runs (O1, O2, O3) and the first part of the fourth observing run (O4a). We also perform comprehensive background simulations to estimate the statistical significance of our top candidates. This is all made possible due to the recently developed strong-lensing analysis pipeline \emph{\textsc{Posterior Overlap 2.0} (PO2.0)}, which has been shown to be near optimal and computationally inexpensive~\citep{barsode2025fast,barsode2026lensing,barsode2026bb}. This pipeline retains the near optimality of the Bayesian model selection by rewriting the Bayesian evidence as an inner product of the posteriors estimated from two individual events that is appropriately weighted by the population prior~\citep{barsode2025fast}. This re-weighted inner product can be computed with very small computational cost, achieving a speed up of $\sim 10^4$ as compared to traditional Bayesian model selection methods~\citep{janquart2021fast, janquart2023return, lo2023bayesian}.

With this search also, we find no evidence for strong lensing in the data: the probability of the top-ranked event being lensed is $< 0.6\%$ (see Fig.~\ref{fig:search_result} and Table~\ref{tab:top_pairs}). We also put a 90\% upper bound on the lensing fraction to be 1.4\%. Five out of the top nine lensed candidate events are from the IAS catalog. We forecast that the probabilities of making a $3 \sigma$ detection by the end of the O4, IR1, and O5 observing runs are $\sim 20\%, \, 23\%,$ and $67\%$ respectively.

This paper is organized as follows: in Sec.~\ref{sec:methodology}, we give a primer on GW strong lensing and describe our methodology for detecting it. Our results are presented in Sec.~\ref{sec:results}, followed by a conclusion in Sec.~\ref{sec:conclusion}. Technical details regarding event selection, prior generation, and foreground-background simulations can be found in the appendices.

\section{Methodology}
\label{sec:methodology}
In the geometric optics regime, the lensed GW polarizations $h^{+,\times}_\mathrm{L}$ are given in terms of the unlensed polarizations $h^{+,\times}_\mathrm{U}$ as
\begin{equation}
\label{eq:generic_lensed_waveform}
h^{+,\times}_\mathrm{L}(f) = \sqrt{\mu} ~ h^{+,\times}_\mathrm{U}(f) ~ e^{\iota (-2\pi f \delta t + n \pi / 2)}
\end{equation}
where $\mu$ is the lensing magnification, $\delta t$ is the time delay, and $n\in\{0,1,2\}$ is a discrete Morse index indicating whether the image forms at the minimum, saddle, or maximum of the time delay surface, respectively~\citep{dai2017waveforms}.

At current detector sensitivities, lensed waveforms are largely degenerate with unlensed ones, as the lensing magnification, time delay and Morse phase shift can be absorbed by a change in the luminosity distance, arrival time, and coalescence phase of the BBH source, all of which are unknown apriori\footnote{in principle, this degeneracy is broken for saddle point images of sources having sufficient contribution from subdominant modes of GW radiation, but this is currently not a significant effect~\citep{dai2020search, vijaykumar2023detection, janquart2021identification}.}. GW strong lensing search can therefore be performed using events that have already been identified using standard BBH search techniques. We outline our event selection criteria in Appendix~\ref{sec:data}.

We model the gravitational lenses as singular isothermal ellipsoids (SIE)~\citep{kormann1994isothermal, fukugita1991gravitational}, which are a good model for galaxy-scale lenses. This lens profile produces 2, 3, or 4 images, some of which may be too dim to be detectable by current detector networks. It is convenient to first search for \emph{pairs} of lensed events, and later aggregate those having common events into a plausible triplet or quadruplet, potentially including a separate search for subthreshold (i.e., relatively dim) counterparts. In this paper, we restrict ourselves to the identification of strongly lensed pairs among superthreshold (i.e., confident) events, and leave the second problem to future work. Thus, we are looking for pairs of events which have similar waveform morphology except for a \textit{relative} magnification $\mu_r$, time delay $\Delta t$, and coalescence phase difference $\dphi/2\in\{0,\pi/4,\pi/2\}$
\begin{equation}
\label{eq:relative_lensed_waveform}
h^{+,\times}_2(f) = \sqrt{\mu_r} ~ h^{+,\times}_1(f) ~ e^{\iota(-2\pi f \Delta t + \dphi)}.
\end{equation}

The optimal statistic to search for such pairs is given by the strong lensing Bayes factor
\begin{equation}
\label{eq:BLU_definition}
\blu = \dfrac{P(d_1, d_2 \mid \HL)}{P(d_1, d_2 \mid \HU)},
\end{equation}
where $ \HL$ is the hypothesis that the two events are lensed copies of the same merger; i.e., they are related by Eq. \eqref{eq:relative_lensed_waveform}, while $\HU$ is the hypothesis that they are unrelated signals. As shown in~\cite{barsode2025fast}, Eq.~\eqref{eq:BLU_definition} can be rewritten as
\begin{multline}
\label{eq:PO2_BLU}
\blu = \frac{1}{\prod\limits_{j=1}^2 P(d_j \mid \HU)} \left[\int d\theq\ d\thbI\ d\dthb ~~ \dfrac{P(\theq, \thbI \mid d_1)}{P_\textsc{PE,1}(\theq, \thbI)} \right.\\
\left. \dfrac{P(\theq, \thbII=\thbI + \dthb \mid d_2)}{P_\textsc{PE,2}(\theq, \thbII=\thbI + \dthb)} ~~ P(\theq, \thbI, \dthb \mid \HL) \right]
\end{multline}
where
\begin{equation}
\label{eq:PO2_ZU}
P(d_j \mid \HU) = \int d\theta\ \dfrac{P(\theta \mid d_j)} {P_{\textsc{PE},j}(\theta)} ~~ P(\theta \mid \HU).
\end{equation}

Above, $\theq$ represents parameters such as masses, spins, sky location, inclination and polarization angles of the BBH that are expected to be identical between two lensed signals, while $\theta_{\mathrm{b}1}$ and $\theta_{\mathrm{b}2}$ are parameters such as the apparent luminosity distance $d_L$, arrival time $t_0$, and coalescence phase $\phi_0$ of the two signals which are expected to be related in the case of lensed signals. Also, $\dthb \equiv \left\{\mu_r=d_{L,1}/d_{L,2}, ~ \dt=t_{0,2}-t_{0,1}, ~ \dphi=\phi_{0,2}-\phi_{0,1}\right\}$, and symbolically, $\thbII=\thbI+\dthb$. The posterior distributions of the BBH parameters obtained from the two events' data $d_1,d_2$ are denoted as $P(\theq, \thbI \mid d_1)$ and $P(\theq, \thbII \mid d_2)$, using $P_\textsc{PE,1}(\theq, \thbI)$ and $P_\textsc{PE,2}(\theq, \thbII)$ as priors. $P(\theq, \thbI, \dthb \mid \HL)$ and $P(\theta \mid \HU)$ are the expected astrophysical priors of detectable lensed and unlensed BBHs.

This reformulation of the Bayes factor assumes that the two signals are well separated, so that the noise is uncorrelated between them. In addition, it assumes that subdominant modes of GW radiation are negligible, as is the case at current detector sensitivities, so that the effect of Morse phase is absorbed in a shift in the coalescence phase. Apart from these two reasonable assumptions, this is mathematically equivalent to doing joint parameter estimation of the two signals.

\subsection{Analysis Prior}\label{sec:priors}

\begin{figure*}[t]
\centering
\includegraphics[width=\columnwidth]{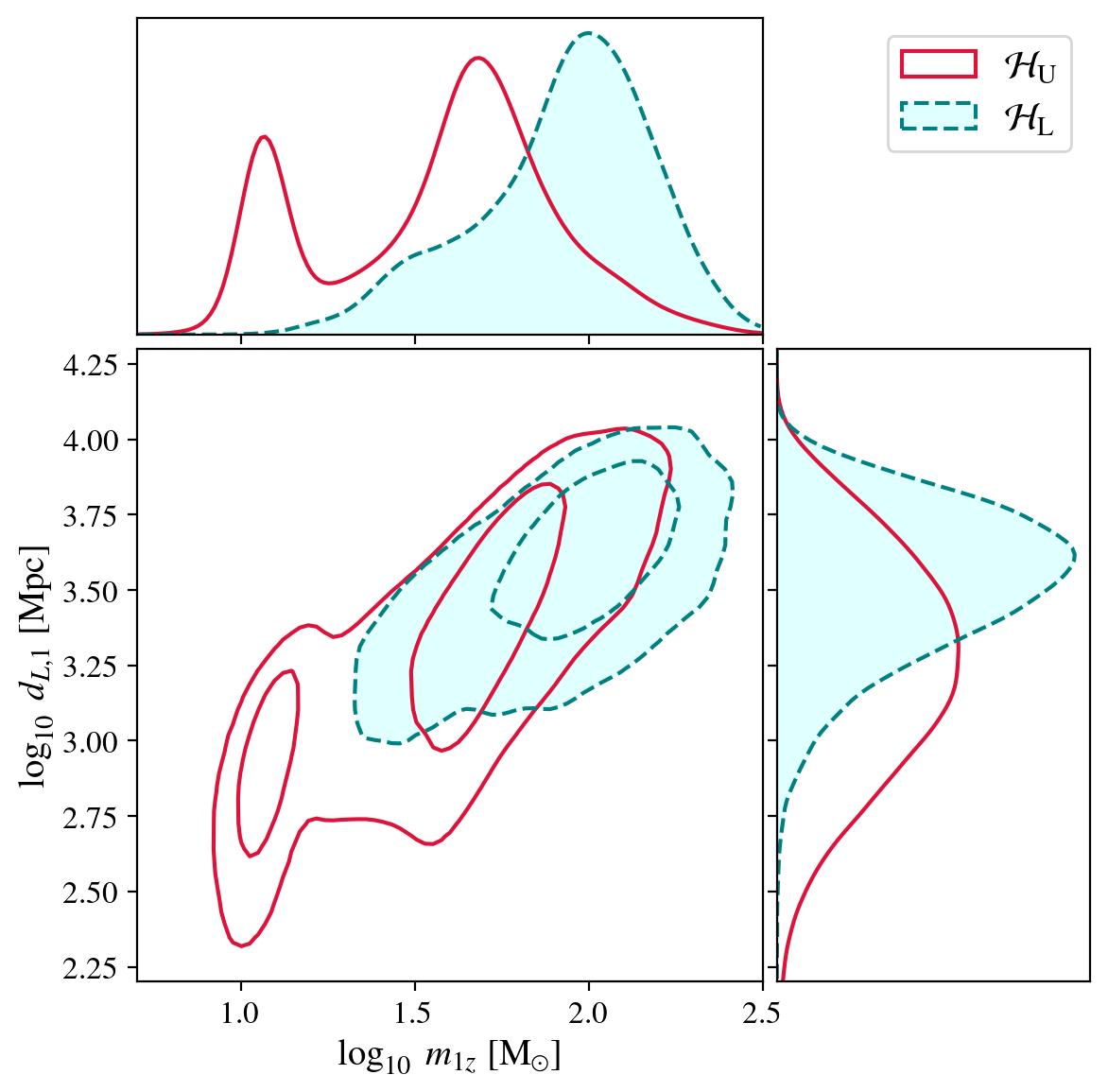}
\includegraphics[width=\columnwidth]{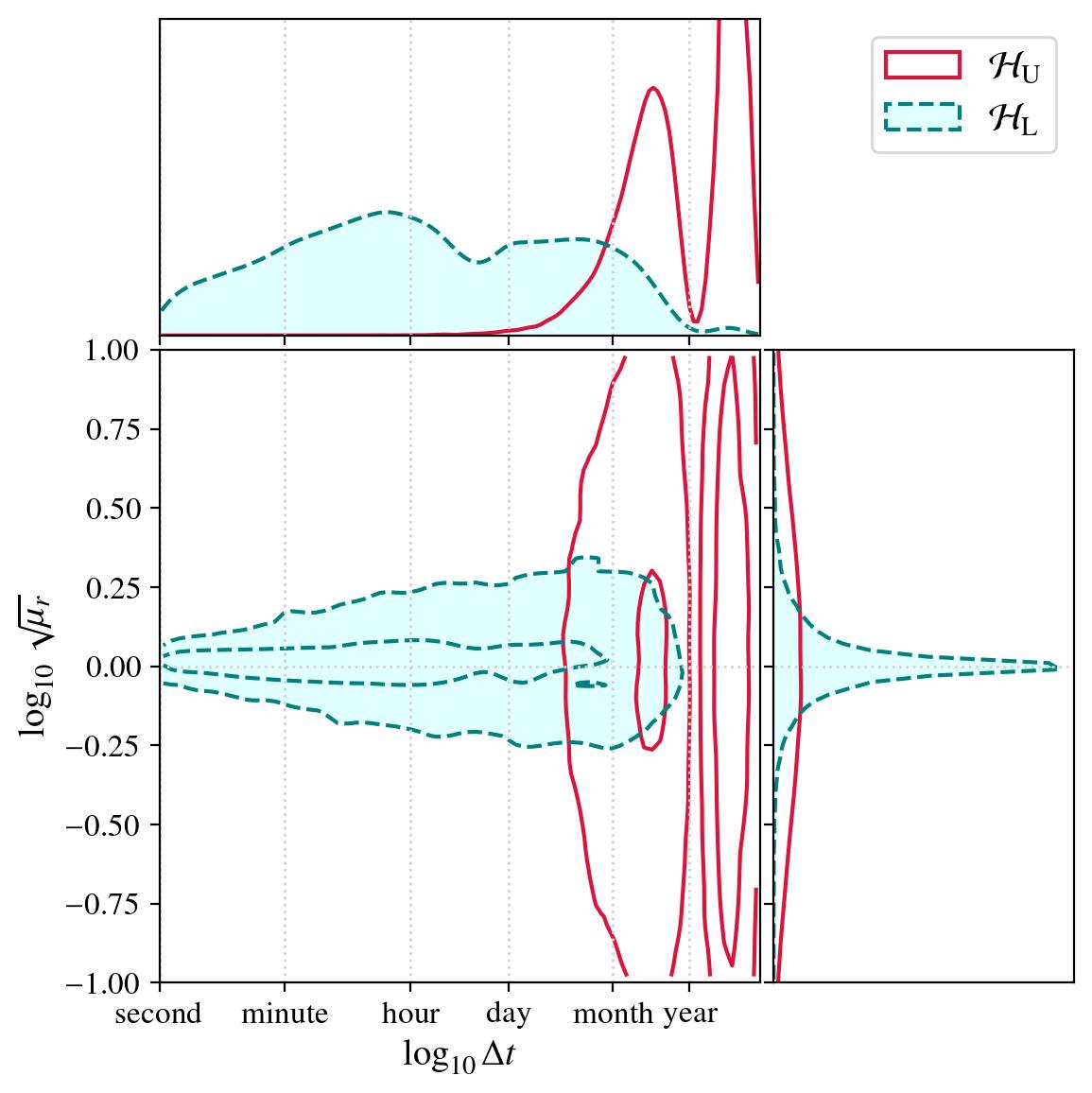}
\caption{\textit{Left}: The prior distributions $P(\log_{10} m_{1z}, \log_{10} d_{L,1} \mid \mathcal{H})$ of the detector frame primary mass and the \textit{apparent} luminosity distance of the first image (or the only image in case of unlensed) under the lensed $(\mathcal{H}=\HL)$ and unlensed $(\mathcal{H}=\HU)$ hypotheses. These are for detectable BBHs. The top and side panels show their marginalized distributions. These distributions are marginalized over all other parameters. \textit{Right}: Same as the left panel, except that the figure shows the prior $P(\log_{10} \dt, \log_{10} \sqrt{\mu_r} \mid \mathcal{H})$ of the lensing time delay and the square root of the relative magnification.}
\label{fig:priors}
\end{figure*}

The Bayes factor---and consequently, the search sensitivity---depends on the data as well as the assumed astrophysical priors $P(\theq, \thbI, \dthb \mid \HL)$ and $P(\theta \mid \HU)$ on the source and lens parameters corresponding to detectable GW signals~\citep{cheung2023mitigating, barsode2025fast}. The source parameter distributions of nearby BBHs are relatively well constrained by the GWTC-4 catalog~\citep{abac2025gwtcpop}, though the high-redshift merger rate remains unconstrained. On the other hand, galaxy surveys, N-body simulations, and the theory of structure formation provide a handle on the population of lenses, though their density profiles are still under active debate~\citep{Shajib_2021, Despali_2026}.

We assume that intrinsic BBH parameters are distributed according to GWTC-4~\citep{abac2025gwtcpop} (we fold in the uncertainties), and extrapolate them to high-redshifts using a merger rate following \cite{dominik2013double}'s isolated binary population synthesis model. We consider a wide range of lens masses between $10^8-10^{15} \mathrm{M}_\odot$, consisting of dark matter halos distributed according to the Halo Mass Function model of~\cite{Behroozi_2013}, and convert them to the velocity dispersion parameter of the SIE lenses following the prescription from \cite{jana2023cosmography, maity2026strong}. The axis ratios of the SIE lenses are assumed to be distributed according to the SDSS catalog of galaxies~\citep{collett2015population}. Selection effects are estimated by rejecting BBHs that have a network optimal S/N $< 8$ or those that do not arrive during the detector on-times reported by the LVK.

Figure~\ref{fig:priors} shows the priors of the detector frame primary mass $m_{1z}$ and the apparent luminosity distance $D$ in the left-hand panel, while the right-hand panel shows the same for the lensing time delay and relative magnification. These are obtained from astrophysical simulations described above and are marginalized over all other parameters.

Note that we have chosen a merger rate distribution that peaks at relatively high redshifts, which results in a higher lensing fraction. Furthermore, our choice of modeling even high-mass dark matter halos (which might host clusters rather than galaxies) with an SIE model also leads to a higher lensing probability than using the Navarro-Frenk-White (NFW) model~\citep{2012arXiv1206.4919S,Brando:2024inp,Vujeva:2025kko}. Since we include high mass halos, it also leads to a larger probability of obtaining higher lensing time delays than the galaxy-only models employed in previous searches, and can account for lensed images arriving in different observing runs. Therefore, our prior can be considered as optimistic towards lensing. The intrinsic lensing fraction is $0.94\%$, which reduces to $0.1\%$ after accounting for selection effects (see Appendix~\ref{sec:popgen}).

\subsection{Determining the significance of lensed events}\label{sec:metrics}

Here, we describe how we draw conclusions on the presence/absence of lensing in the data. These metrics are derived from the Bayes factors after including additional information from the prior and background-foreground analyses.

We empirically find that the PO2.0 Bayes factor is reasonably accurate even in real noise (see Appendix~\ref{sec:bgfg}), which allows us to define and interpret the relative posterior odds between the lensed and unlensed hypotheses
\begin{eqnarray}
\label{eq:posterior_odds}
\olu \equiv \dfrac{P(\HL \mid d_1, d_2)}{P(\HU \mid d_1, d_2)} = \blu ~ \plu,
\end{eqnarray}
where the prior odds $\plu\equiv P(\HL)/P(\HU)$ of finding a strongly lensed GW in a catalog of $N$ signals are $\approx 2u/(N-1)$~\citep{hannuksela2025strong}. $\olu \gtrsim 1$ would indicate evidence in favor of the lensed hypothesis, though the uncertainty in our prior knowledge may cause unknown systematic errors.

We can estimate the frequentist error rates using a simulated unlensed background and a lensed foreground (details in Appendix~\ref{sec:bgfg}). Using the distributions of $\blu$ from these simulations and accounting for the size of the catalog, the false positive probability ($\fpp$) that at least one unlensed pair appears as an outlier is given by
\begin{equation}
\label{eq:FPPcat}
\fpp(\blu) \equiv 1 - \mathrm{cdf}(\blu \mid \HU)^{N(N-1)/2},
\end{equation}
while the false dismissal probability ($\fdp$) of discarding a true lensed pair is given by
\begin{equation}
\label{eq:FDP}
\fdp(\blu) \equiv \mathrm{cdf}(\blu \mid \HL).
\end{equation}
$\fpp \lesssim 10^{-6}$ would be a decisive rejection of the unlensed hypothesis and a detection of lensing, while $\fpp \lesssim 0.05$ may be sufficient to prompt follow-up. The $\fdp$ tells us about the efficiency of our pipeline at identifying true lensed events, with smaller values indicating better performance.

Given the simulated distributions of $\blu$ from the lensed and unlensed hypotheses, we can also define the probability that a given pair of events is lensed (this is similar to the probability of astrophysical origin $p_\mathrm{astro}$ that is commonly used in GW searches~\citep{abbott2016binary})
\begin{equation}
\label{eq:plensed}
\plensed(\blu) \equiv \left[ 1 + \dfrac{1}{\plu} \dfrac{P(\blu \mid \HU)}{P(\blu \mid \HL)} \right]^{-1}.
\end{equation}
Note that this would be equal to $P(\HL \mid d_1, d_2) = \olu / (1 + \olu)$ if all our modeling assumptions regarding priors, signal morphology, and noise matched reality (see \cite{barsode2026bb} and also Appendix~\ref{sec:bgfg}). $\plensed \gtrsim 0.5$ may be considered sufficient evidence in favor of the lensed hypothesis.

If we were to consider all pairs with strong lensing Bayes factor higher than $\blu$ as lensed candidates, a catalog of such pairs will have the following purity (i.e., the fraction of truly lensed pairs)
\begin{equation}
\label{eq:purity}
\purity(\blu) \equiv \left[ 1 + \dfrac{1}{\plu} \dfrac{1 - \mathrm{cdf}(\blu \mid \HU)}{1 - \mathrm{cdf}(\blu \mid \HL)} \right]^{-1}.
\end{equation}
Note that, unlike previous metrics that deal with the statistical significance of a given pair, the catalog purity quantifies the detection/non-detection at a population level. The ultimate goal of a strong lensing search is to produce high-purity lensed catalogs for enabling downstream science. $\purity \gtrsim 0.5$ may be sufficient for many studies based on a population of lensed detections~\citep{jana2024strong}, while others may demand higher purity (say, $\purity \gtrsim 0.9$).

\section{Results}
\label{sec:results}

\begin{figure*}[t]
\centering
\includegraphics[width=1.8\columnwidth]{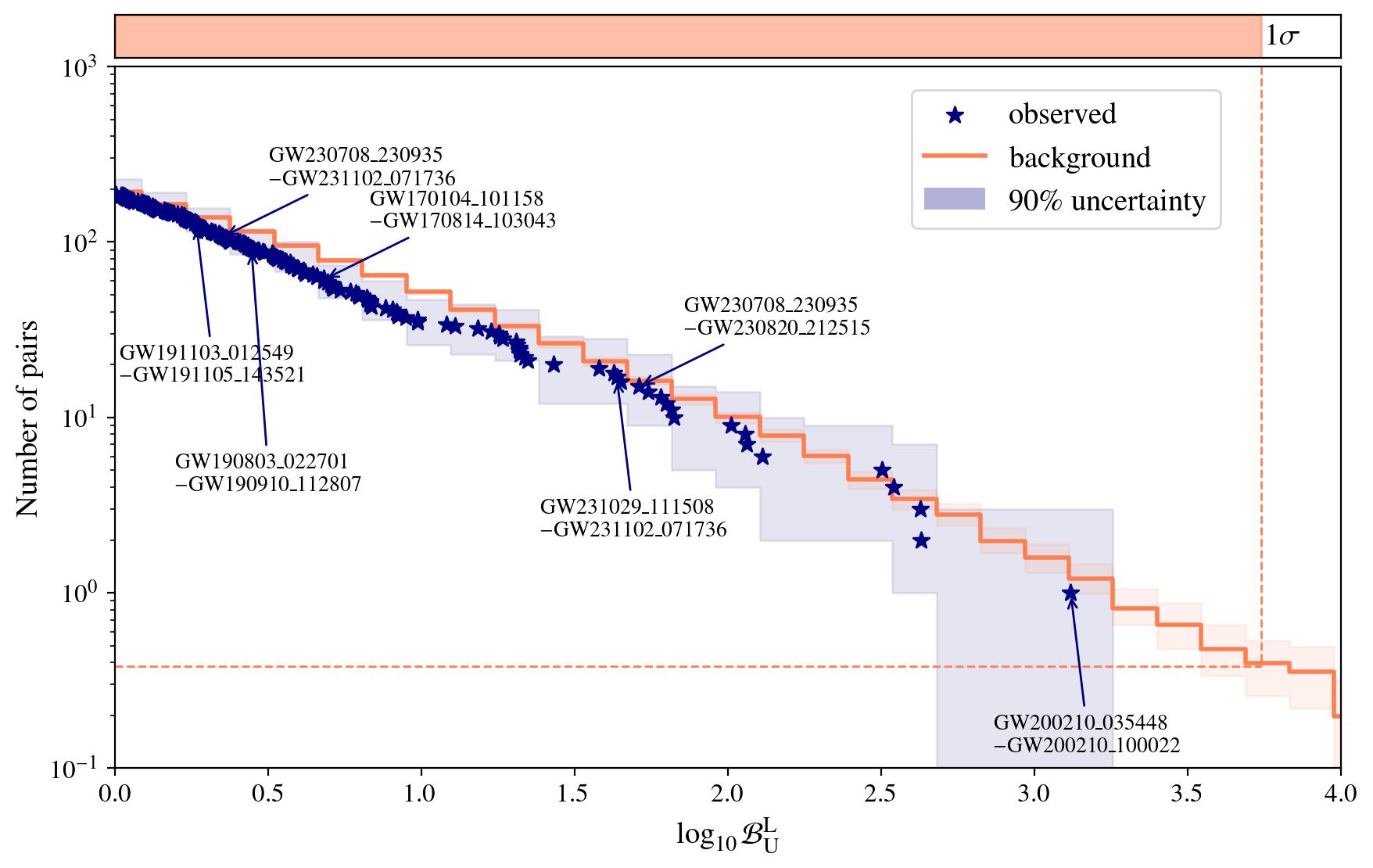}
\caption{The number of pairs above $\log_{10} \blu$ obtained from GW events detected in O1-O4a. Also shown is the background distribution obtained from pairs of unlensed signals injected in real detector noise. Our top pair, as well as several interesting pairs identified in previous searches (if they cross $\blu>1$ in our search), are annotated.}
\label{fig:search_result}
\end{figure*}

We compute the $\blu$ (Eq.~\eqref{eq:PO2_BLU}) on a total of 21321 pairs formed out of 207 unique BBH events in the LVK, OGC, and IAS catalogs. Figure~\ref{fig:search_result} shows the number of pairs found to have $\blu$ above a threshold for the prior described in Sec.~\ref{sec:priors}. For comparison, we also compute $\blu$'s on background and foreground distributions under the same prior, and plot the average number of unlensed pairs expected to lie above the same $\blu$ thresholds. We find no potentially lensed outliers.

We quantify this non-detection below by showing that none of our top pairs, or any of the pairs identified in previous searches, show a high consistency with the lensing hypothesis under the metrics described in Sec.~\ref{sec:metrics}.

\subsection{Top Candidate Pairs}
\begin{deluxetable*}{l c c c c c c}
\label{tab:top_pairs}
\tablecaption{For the top 5 pairs in our search, the pair's Bayes factor ($\blu$, Eq.~\eqref{eq:PO2_BLU}), posterior odds ($\olu$, Eq.~\eqref{eq:posterior_odds}), false positive probability ($\fpp$, Eq.~\eqref{eq:FPPcat}), false dismissal probability ($\fdp$, Eq.~\eqref{eq:FDP}), probability of being lensed ($\plensed$, Eq.~\eqref{eq:plensed}), and the purity of a catalog that would contain pairs at least as significant as these ($\purity$, Eq.~\eqref{eq:purity}).}
\tablehead{\colhead{Pair} & \colhead{$\log_{10} \blu$} & \colhead{$\log_{10} \olu$} & \colhead{$\fpp$} & \colhead{$\fdp$} & \colhead{$\plensed \times 10^3$} & \colhead{$\purity \times 10^2$}}
\startdata
GW200210\_035448\tablenotemark{*} --~GW200210\_100022\tablenotemark{*} & 3.12 & -1.89 & 0.69 & 0.27 & 5.65 & 11.47  \\
GW190425\_133124\tablenotemark{*} --~GW190605\_025957\tablenotemark{*} & 2.63 & -2.38 & 0.95 & 0.21 & 2.28 & 5.20  \\
GW230707\_124047 --~GW230708\_230935 & 2.63 & -2.39 & 0.95 & 0.21 & 2.27 & 5.17  \\
GW230922\_040658 --~GW231005\_021030 & 2.54 & -2.47 & 0.97 & 0.21 & 1.93 & 4.56  \\
GW190425\_133124\tablenotemark{*} --~GW190426\_190642 & 2.50 & -2.51 & 0.97 & 0.21 & 1.80 & 4.39  \\
\enddata
\tablenotetext{*}{these events were first reported by the IAS pipeline}
\end{deluxetable*}
Table~\ref{tab:top_pairs} shows the inference metrics for the 5 top-ranked pairs under the assumed prior. Among all pairs, the maximum posterior odds $\olu = 10^{-1.89}$ are too small, while the minimum $\fpp = 0.69$ is too large, to be considered as lensed. The probability that any of these pairs are lensed is $< 0.6\%$, and if we were to construct a catalog out of pairs like these, it would be at most $12\%$ pure (i.e., $\sim7$ out of 8 times it would just contain unlensed pairs).

The top pair, GW200210\_035448--GW200210\_100022, does not show a high amount of consistency in the inferred parameters. However, their high (total mass $\sim 200 \mathrm{M}_\odot$), asymmetric (mass ratio $\sim$1/3) detector frame masses, and a time delay of $\sim 6$ hours are highly favored by the lensed prior over the unlensed prior (see Fig.~\ref{fig:priors}). All of our top 5 pairs appear to be favored by the prior, as they all have moderate to high detector frame total masses $120-450 \mathrm{M}_\odot$, with mass ratios ranging from $0.3-1$, and time delays of hours to days.

Notably, among the 10 unique events in the top 5 pairs, 5 are from the IAS catalog. This is likely due to the fact that high-mass events are favored by the lensing prior, and the IAS pipeline is more sensitive in those regions due to more powerful ranking statistics and the inclusion of subdominant modes of GW radiation.

\subsection{Previously Identified Candidate Pairs}

\begin{deluxetable*}{l l c c c c c c c}
\label{tab:prev_pairs}
\tablecaption{For the candidate pairs identified in previous searches, the pair's rank, Bayes factor ($\blu$, Eq.~\eqref{eq:PO2_BLU}), posterior odds ($\olu$, Eq.~\eqref{eq:posterior_odds}), false positive probability ($\fpp$, Eq.~\eqref{eq:FPPcat}), false dismissal probability ($\fdp$, Eq.~\eqref{eq:FDP}), probability of being lensed ($\plensed$, Eq.~\eqref{eq:plensed}), and the purity of a catalog that would contain pairs at least as significant as these ($\purity$, Eq.~\eqref{eq:purity}).}

\tablehead{\colhead{Pair} & \colhead{First identified by} & \colhead{Rank} & \colhead{$\log_{10} \blu$} & \colhead{$\log_{10} \olu$} & \colhead{$\fpp$} & \colhead{$\fdp$} & \colhead{$\plensed \times 10^3$} & \colhead{$\purity \times 10^2$}}

\startdata
GW170809\_082821 --~GW170814\_103043 & \cite{hannuksela2019search} & 284 & -0.28 & -5.30 & 1.00 & 0.02 & 0.02 & 0.08  \\
GW170104\_101158 --~GW170814\_103043 & \cite{hannuksela2019search} & 61 & 0.68 & -4.33 & 1.00 & 0.06 & 0.09 & 0.25  \\
GW190803\_022701 --~GW190910\_112807 & \cite{LIGOScientific:2021izm} & 92 & 0.45 & -4.57 & 1.00 & 0.05 & 0.06 & 0.19  \\
GW190620\_030421 --~GW200216\_220804 & \cite{abbott2023search} & 537 & -0.86 & -5.87 & 1.00 & 0.02 & 0.01 & 0.05  \\
GW191103\_012549 --~GW191105\_143521 & \cite{janquart2023follow} & 128 & 0.27 & -4.75 & 1.00 & 0.04 & 0.05 & 0.15  \\
GW230708\_230935 --~GW230820\_212515 & \cite{abac2025gwtclens} & 15 & 1.71 & -3.30 & 1.00 & 0.12 & 0.45 & 1.17  \\
GW230708\_230935 --~GW231102\_071736 & \cite{abac2025gwtclens} & 107 & 0.35 & -4.66 & 1.00 & 0.04 & 0.05 & 0.17  \\
GW231029\_111508 --~GW231102\_071736 & \cite{abac2025gwtclens} & 17 & 1.64 & -3.38 & 1.00 & 0.12 & 0.40 & 1.06  \\
\enddata
\end{deluxetable*}

Table~\ref{tab:prev_pairs} shows our inference metrics for pairs of events that were identified in previous searches as interesting or the most significant, though none were claimed to be truly lensed. We find the same result: all of these pairs are ranked 15 or above in our search, exhibiting low scores on all of our metrics, and thus are most likely unlensed.

There is an interesting pattern in the candidate pairs. Barring the exception of GW191103\_012549--GW191105\_143521 (which was vetoed in the main search by \cite{abbott2023search}, but was later reanalyzed in \cite{janquart2023follow}), candidates have generally gotten heavier chronologically, with total detector frame masses of $\sim 80 \mathrm{M}_\odot$ for \cite{hannuksela2019search}'s top pair GW170809\_082821--GW170814\_103043, to $\sim 160 \mathrm{M}_\odot$ for the top 3 pairs identified in the latest search by the LVK~\citep{abac2025gwtclens}, over to even more massive $\gtrsim 200 \mathrm{M}_\odot$ candidates identified in our search.

This pattern is likely caused by an effect discussed by \cite{cheung2023mitigating}: if a region of the parameter space (here, the mass) is favored by (detectable) unlensed events, we are more likely to encounter false positives there. Initial searches did not fold this into their statistic, and consequently recovered candidates mainly in the peak sensitivity band of the detectors. Later searches included population models in the Bayes factor calculation, reducing the weight in the favored regions, which increased the same in the disfavored regions of the unlensed population. Therefore, the candidates---which are most likely unlensed pairs---were identified near the fringes of the detectable population, i.e., in the low-probability, high-mass end. As the detector sensitivity has improved over the years, this region has shifted towards ever higher masses.

\subsection{Constraints on the Strong Lensing Fraction}

\begin{figure}[t]
\centering
\includegraphics[width=\columnwidth]{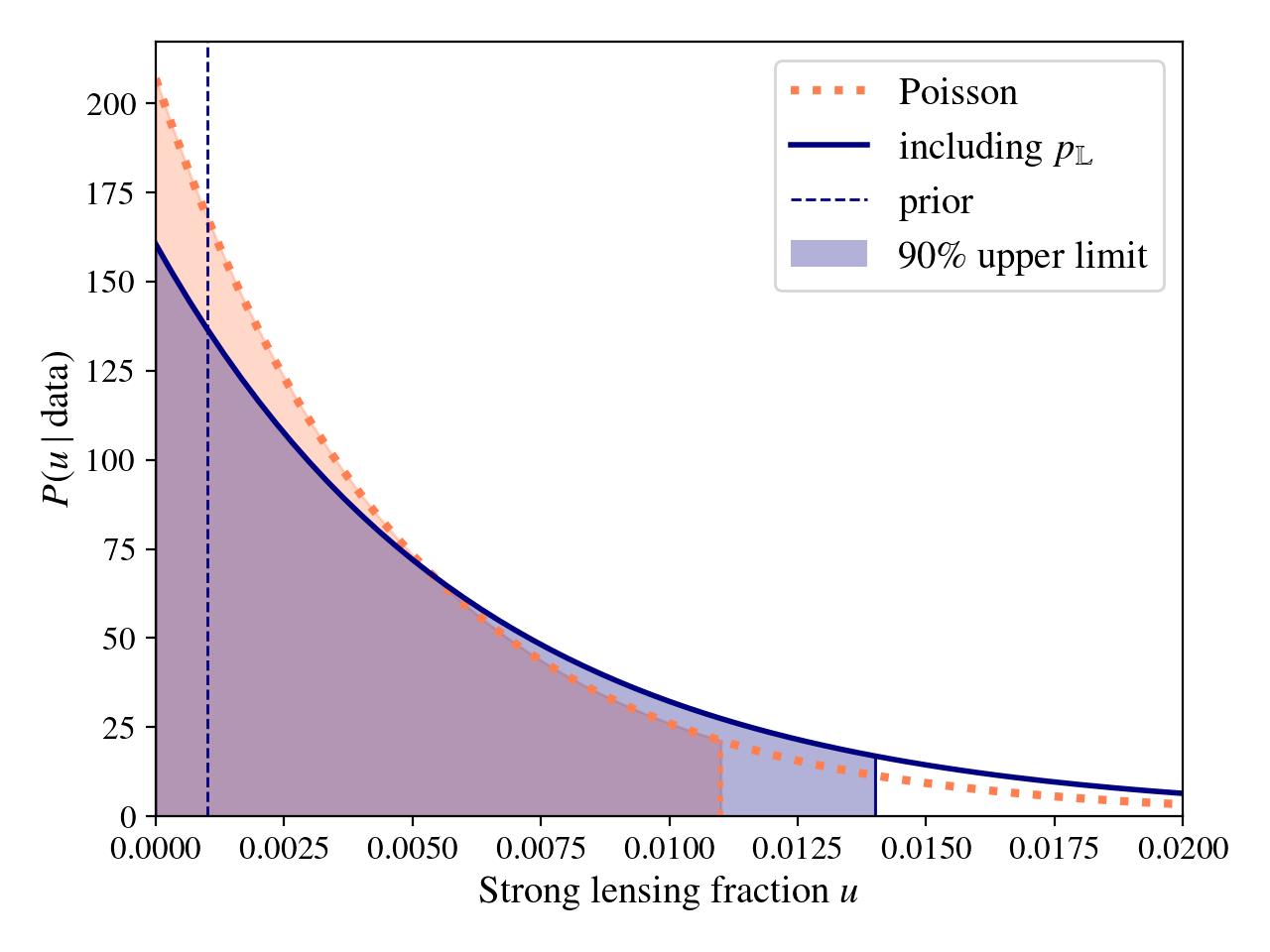}
\caption{Constraints on the strong lensing fraction obtained using the non-observation in all events detected in O1-O4a. The orange dotted curve shows the constraints obtained using Poisson statistics, assuming that all 207 events are unlensed with complete certainty, while blue and cyan curves consistently account for the very small, but non-zero probability that some of the pairs may be lensed.}
\label{fig:u_posteriors}
\end{figure}

Since none of the observed pairs show evidence in favor of lensing, we constrain the fraction $u \equiv {\Lambda_\ell}/{\Lambda}$ of GWs that may be strongly lensed, where $\Lambda_\ell$ is the rate of strongly lensed pairs (Poisson mean), and $\Lambda$ is the total rate of unlensed events and lensed pairs. In previous work~\citep{basak2022constraints}, this was achieved by assuming that all of the events are unlensed with complete certainty. Then the posterior on $u$ can be obtained assuming Poisson statistics (Eq.~7 of \cite{basak2022constraints})
\begin{equation}
\label{eq:u_posterior_Poisson}
P(u) \propto \int\limits_{0}^{\infty} \Lambda^{N+1} ~ e^{-(1+u)\Lambda} ~ P_\ell(u \Lambda) P_\Lambda(\Lambda) d\Lambda.
\end{equation}
This is essentially a marginalization over $\Lambda$ assuming priors $P_\ell$ and $P_\Lambda$ on the lensed and unlensed rates. We assume these priors to be uniform and fix the normalization numerically.

This analysis has the drawback that it forces us to assign discrete labels (certainly unlensed or certainly lensed) to each event, and cannot take into account our uncertainty in making such an identification for each pair. Therefore, we adopt a different approach recently put forth by \cite{harshe2026could}, where we consider the observed $\blu$ distribution to be a mixture of the foreground and background $\blu$ distributions, with the mixture coefficient equal to the prior odds $\plu(u)\approx 2u/(N-1)$:
\begin{multline}
\label{eq:u_posterior_mixture}
P(\{\blu\} \mid u) = \prod\limits_{i=1}^{N(N-1)/2} \left[ \plu(u) ~ P({\blu}_i \mid \HL) \right. \\
\left. + ~ (1 - \plu(u)) ~ P({\blu}_i \mid \HU) \right].
\end{multline}
This approach consistently includes our uncertainty in determining the exact nature (lensed or unlensed) of the event pairs, and will work even in the post-detection era.

Assuming uniform priors on $u$, we plot in Fig.~\ref{fig:u_posteriors} the posterior distributions of $u$ obtained using the simple Poisson and the mixture approach. The Poisson posteriors are the tightest, while the mixture model yields slightly less constraining results since it considers the possibility that some of the pairs may be lensed. Based on these, we put a $90\%$ upper bound on the lensing fraction at $\sim 1.4\%$.

\section{Conclusion}
\label{sec:conclusion}

\begin{figure}[t]
\centering
\includegraphics[width=\columnwidth]{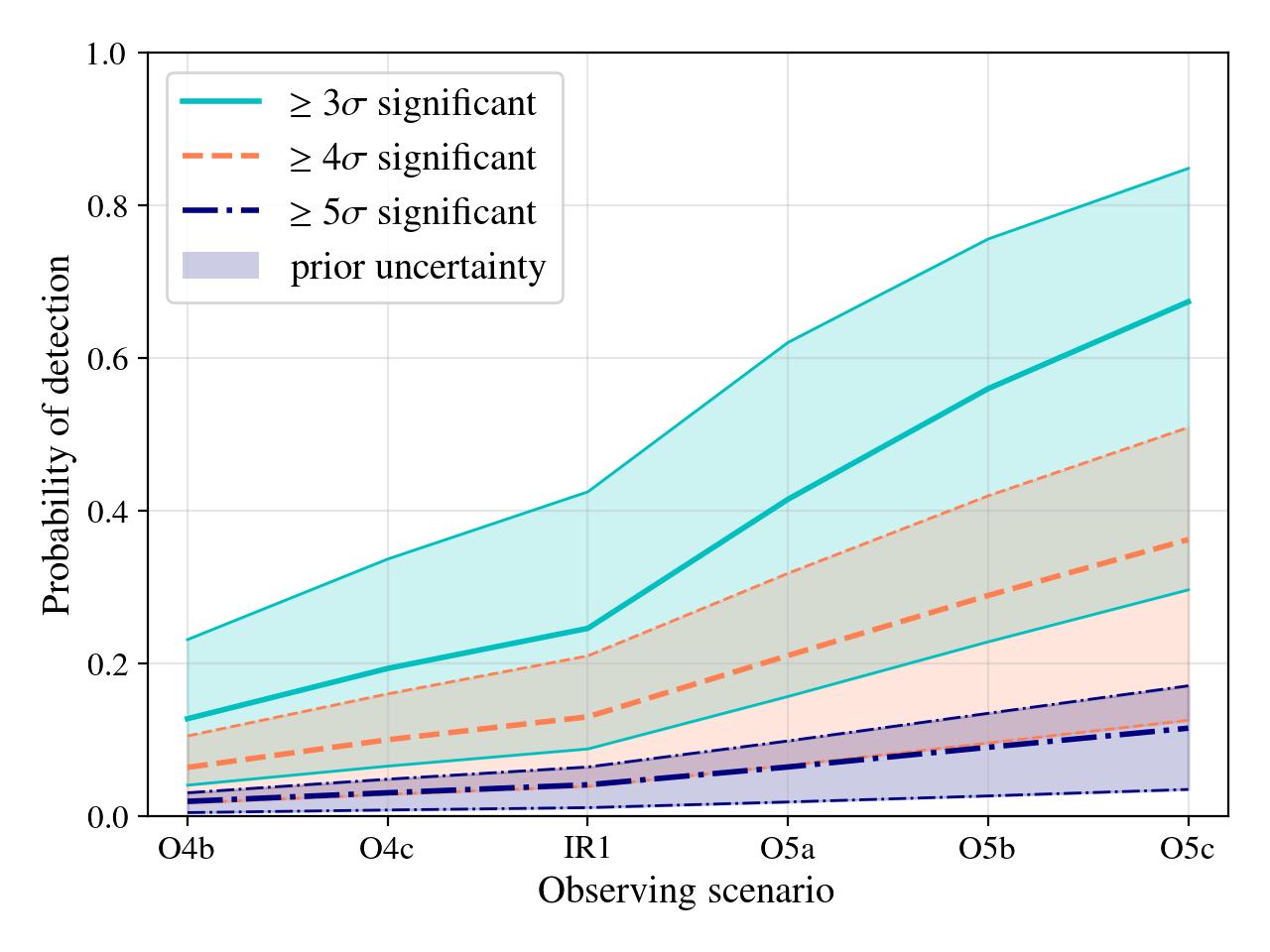}
\caption{The probability of finding at least one strongly lensed pair of GWs in the data at the end of various observing run segments. Results are shown for different thresholds on the statistical significance, with shaded regions indicating uncertainties due to those in our prior knowledge (see Appendix~\ref{sec:forecast} for details).}
\label{fig:forecast}
\end{figure}

We have searched for signatures of strongly lensed GWs among the BBH events reported in the LVK, IAS, and OGC catalogs. Based on realistic background and foreground simulations, we estimated the probability of lensing as a new metric to assess the statistical significance of a lensed candidate pair, and the purity of a candidate catalog as a means to assess the same at a population level. Even with a prior favoring the lensing hypothesis, we found nothing statistically significant, and have thus put a 90\% upper limit on the strong lensing fraction at 1.4\%.

The inclusion of events from the IAS catalog marks a significant improvement over existing searches. Due to their higher sensitivity at the high-mass range of the compact binary parameter space, the IAS catalog recovers more massive events, which are in turn more easily identifiable, if lensed. However, a caveat here is that the combined population of observed GWs no longer follows the selection criteria described by either one of the search pipelines. This may have biased our prior population to some degree. Correcting this bias would require us to consider selection effects beyond the simple signal-to-noise ratio (S/N) based cut adopted here, and is left as future work.

Our results are conditioned on our particular choice of prior. While we have chosen to be optimistic towards lensing, the rate of high-redshift GW transient sources, as well as the population and properties of the lenses themselves, are largely unconstrained. A comprehensive search should fold in these uncertainties, though, at the moment, it is too computationally expensive to do so for more than a few discrete choices of the prior. We are currently exploring ways to search with a continuum of prior choices through efficient reweighting of existing results, to be included in future searches.

Through this work, we demonstrated the ability of PO2.0 to perform a low-cost search for lensed GWs at near-optimal efficiency, in real detector noise. In the future, we will continue to search in the latest available data for signatures of strongly lensed GWs. Our forecast for when we might see such signatures is presented in Fig.~\ref{fig:forecast}. While a $\gtrsim 5 \sigma$ detection may have to wait, a $\gtrsim 3 \sigma$ detection may likely occur (with $> 50\%$ probability) sometime within the next three years, i.e., during the fifth observing run of the LVK. Optimistically, there is a 40\% chance even by the end of the intermediate IR1 run, i.e., within the next one year. Exciting times are ahead for kickstarting astrophysics and cosmology using strongly lensed GWs.

\section*{Acknowledgments}
We thank Jef Heynen for a careful review of this manuscript. We also thank Tejaswi Venumadhav, Srashti Goyal, and members of the ICTS Astrophysics \& Relativity group and the LVK Lensing group for constructive feedback on this work. We acknowledge the support of the Department of Atomic Energy, Government of India, under project nos. RTI4019 and RTI4013. The numerical calculations reported in the paper were performed on the Alice computing cluster at ICTS-TIFR.

This work makes use of the \texttt{cogwheel} \citep{roulet2022removing, islam2022factorized}, \texttt{scipy} \citep{2020SciPy-NMeth}, \texttt{bilby} \citep{bilby_paper}, and \texttt{pyCBC} \citep{alex_nitz_2024_10473621} software packages.

This research has made use of data or software obtained from the Gravitational Wave Open Science Center (gwosc.org), a service of the LIGO Scientific Collaboration, the Virgo Collaboration, and KAGRA. This material is based upon work supported by NSF's LIGO Laboratory, which is a major facility fully funded by the National Science Foundation, as well as the Science and Technology Facilities Council (STFC) of the United Kingdom, the Max-Planck-Society (MPS), and the State of Niedersachsen/Germany for support of the construction of Advanced LIGO and construction and operation of the GEO600 detector. Additional support for Advanced LIGO was provided by the Australian Research Council. Virgo is funded, through the European Gravitational Observatory (EGO), by the French Centre National de Recherche Scientifique (CNRS), the Italian Istituto Nazionale di Fisica Nucleare (INFN), and the Dutch Nikhef, with contributions by institutions from Belgium, Germany, Greece, Hungary, Ireland, Japan, Monaco, Poland, Portugal, and Spain. KAGRA is supported by the Ministry of Education, Culture, Sports, Science and Technology (MEXT), Japan Society for the Promotion of Science (JSPS) in Japan; National Research Foundation (NRF) and Ministry of Science and ICT (MSIT) in Korea; Academia Sinica (AS) and National Science and Technology Council (NSTC) in Taiwan.

\appendix

\section{GW Events and Data}
\label{sec:data}
We obtain a list of all GW events confirmed by the LVK to have a probability of astrophysical origin $p_\mathrm{astro} > 0.5$ from
\textsc{GWOSC}~\citep{abbott2021open, abbott2023open, abac2025open}, following instructions from the Gravitational Wave Open Data Workshops (\texttt{https://gwosc.org/workshops/}). For the IAS catalog, these are manually added from their publications for the O2~\citep{venumadhav2020new}, O3a~\citep{olsen2022new}, and O3b searches~\citep{mehta2025new}, and read from the GitHub repository linked in \cite{wadekar2023new} for the search including subdominant modes of GW radiation. We read the latest OGC catalog from the GitHub repository linked in \cite{nitz20234}.

We remove events with a 90\% lower bound on the secondary mass $<3 \mathrm{M}_\odot$ to formally restrict to BBHs. Among the LVK events, we further restrict to events with FAR $<$ 1 per year, for which parameter estimation was performed. Our final list contains 207 BBH events, of which 10 are unique to OGC, and 28 are unique to IAS.

We download the LVK events' parameter estimation samples from \textsc{GWOSC}, and the OGC and IAS events’ from their respective GitHub repositories. If, for an event, samples are available from multiple catalogs, we give preference to LVK, IAS, and OGC in that order. We ensure all the posterior samples follow the same priors, reweighting appropriately whenever needed. For weighted samples, we work with a redrawn, equally weighted sample set.

For some OGC events, the coalescence phase posterior is unavailable, while the polarization angle posterior is absent for all of them. We fill these parameters with a uniform distribution between their physically valid ranges. This may result in Bayes factors smaller by at most a factor of $\sim 11$ for pairs involving these events, as the information in the plane of these two parameters gets washed out. However, the maximum measured $\blu$ for any pair involving an OGC event is $\sim 20$. Thus, it will not make it to our list of top 5 pairs (all of which have $\blu > 316$) even if it gets boosted by a factor of $11$, and our overall conclusions will not be affected.

Some of the earlier IAS posterior samples were obtained with a dominant-mode, aligned-spin-only analysis. We do not expect this to affect our conclusions significantly due to the low impact of subdominant modes and the generally poor measurability of spin precession angles at those early detector sensitivities.

\section{Prior Generation}
\label{sec:popgen}

\begin{figure*}[t]
\centering
\includegraphics[width=2\columnwidth]{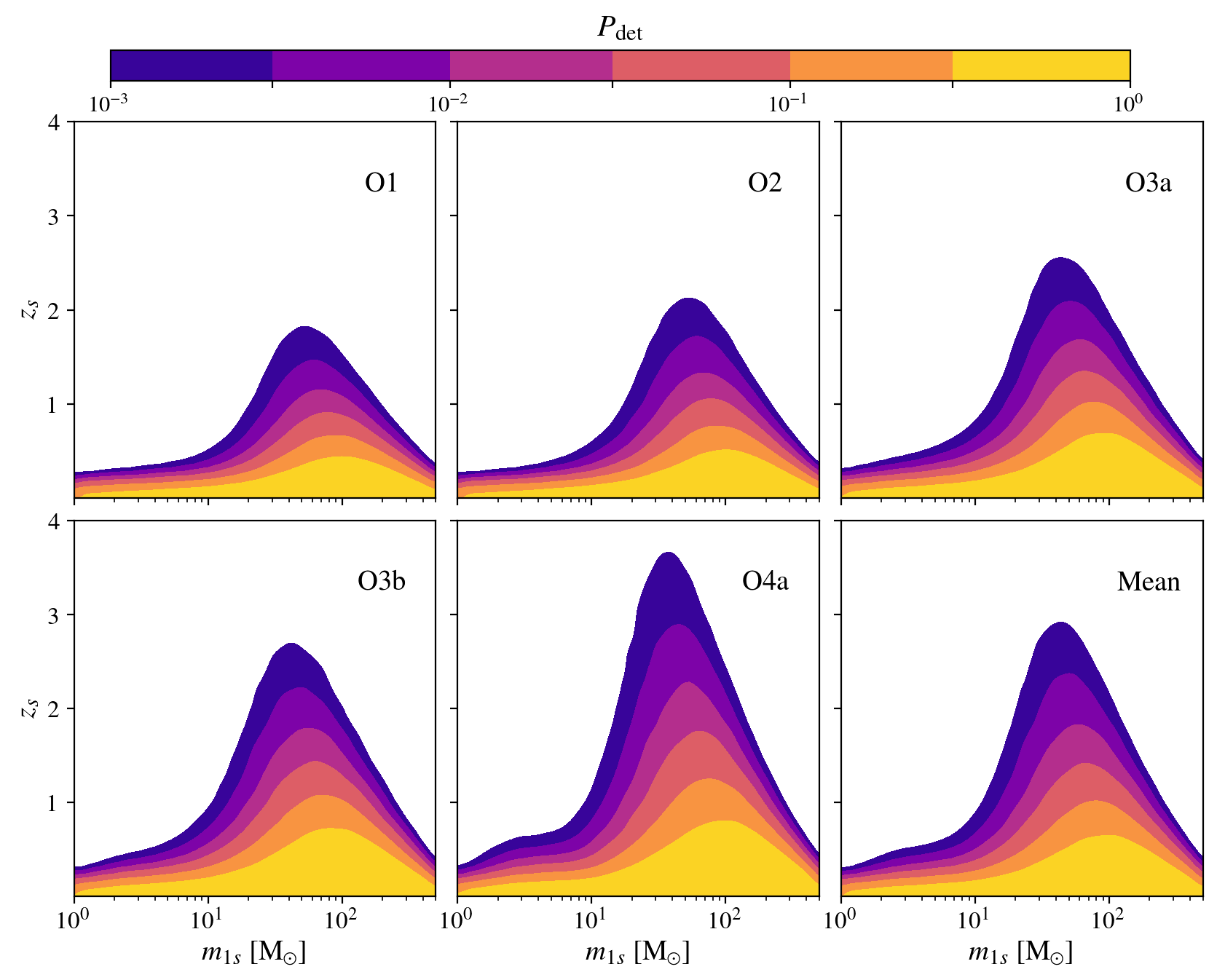}
\caption{The approximate probability $\pdet(m_{1z},z_s)$ that a BBH merger with source frame primary mass $m_{1s}$ at a redshift $z_s$ will be detectable in O1-O4a observing runs. These are marginalized over the remaining BBH parameters, assuming they are distributed according to the GWTC-4 population. The bottom right panel shows the same averaged over all detector on-times during O1-O4a.}
\label{fig:approx_Pdet}
\end{figure*}

In the context of GW lensing search with PO2.0, the prior consists of samples of detectable unlensed events and lensed pairs. Our overall scheme for generating the prior is the same as \cite{haris2018identifying}'s. We sample source BBH parameters and redshifts as described in Sec.~\ref{sec:priors} of the main text\footnote{Our population generation code ignores correlations between the hyperposteriors of different inferred parameters. Later tests showed the resulting bias to be negligible compared to the statistical uncertainty for our chosen number of injections.}, while their arrival times and remaining extrinsic parameters are drawn uniformly and isotropically. Depending on the lensing optical depth to their redshift, sources are randomly assigned to be lensed or unlensed. If they are deemed lensed, lens parameters are drawn according to the differential optical depth. We choose the same cosmological parameters as GWTC-4~\citep{abac2025gwtcpop}.

We impose two kinds of selection effects: we consistently incorporate the duty cycle of each detector by enforcing the BBHs' arrival times to lie during the time at which at least one detector was operational. Secondly, we require their optimal S/N, computed with the \textsc{IMRPhenomXPHM}~\citep{pratten2021computationally} waveform model, to be greater than 8. The noise power spectral densities (PSDs) in each segment are computed by taking the median of ten Welch estimates made randomly within that segment's duration using \textsc{GWPy}. S/N are computed using only the detectors that were operational at the events' arrival times. Unlensed BBH samples that satisfy the above selection criteria are deemed detectable. For lensed events, all possible pairs of images in which both images satisfy the above selection criteria are deemed detectable.

The above procedure is rather simple, but highly computationally expensive in practice. The main bottleneck is that the sampling efficiency (the fraction of unlensed samples drawn from the astrophysical population that has S/N $>$ 8) is only about 1 in $10^4$ at O1-O4a sensitivities. For accurate computation of the $\blu$, we need $\sim 5\times 10^4$ samples, which would require $\sim 5\times 10^8$ S/N calculations in the brute force approach. Worse, the sampling efficiency is an order of magnitude lower for lensed events.

One solution is to compute the S/N using fast techniques such as neural network surrogates or approximations~\citep{phurailatpam2024gwsnr, phurailatpam2024ler}. However, apart from small unmodeled errors in these approximations, there is a significant cost involved in training the machine.

Thus, instead of accelerating S/N calculation, we focus on improving the sampling efficiency using an importance sampling approach. We estimate an approximate probability of detection $\pdet$, using which we make a pre-selection of BBHs that are likely to be detectable. Only for these BBHs, we compute the exact S/N, and keep samples crossing 8. To compensate for the pre-selection, each sample is weighted by $1/\pdet$, and a final equally-weighted sample set is generated using multinomial resampling.

To estimate the approximate $\pdet$, we split the BBH parameters into two sets: $\vec{\theta}$ consisting of luminosity distance, chirp mass, mass ratio, spin magnitudes, spin tilt angles, and inclination angle; and $\vec{\phi}$ consisting of the rest. We then model the full $\mathrm{S/N}(\theta, \phi)$ by the following
\begin{equation}
\label{eq:mean_SNR}
\mathrm{S/N}(\theta, \phi) \sim \mathcal{T}_\nu(\mathrm{\overline{S/N}}(\theta), \sigma, \nu=10)
\end{equation}
i.e., the exact S/N as a function of $\theta, \phi$ is distributed as a Student's $t$ distribution, $\mathcal{T}_\nu$, around a mean S/N as a function of $\theta$, with standard deviation $\sigma$, and the tail parameter $\nu=10$. For a BBH sample with parameters $\theta,\phi$, the approximate $\pdet$ is taken to be the (reversed) cumulative distribution of $\mathcal{T}_\nu$ evaluated at the S/N threshold, which we choose to be slightly smaller than 8 ($\sim 6$) for safety. We train a small $16 \times 8 \times 4$ fully-connected neural network to learn $\mathrm{\overline{S/N}}(\theta)$ and $\sigma$. The resulting $\pdet$ for different observing runs is shown in Fig.~\ref{fig:approx_Pdet}.

Equation~\eqref{eq:mean_SNR} is rather crude, and may not represent the true behavior of S/N. However, it yields a $\pdet$ that is good enough for importance sampling. We can monitor the overall accuracy of this method by ensuring that the reweighting efficiency is not too small: i.e., $\eta_\mathrm{rw} = {\langle 1/\pdet \vphantom{1/\pdet^2} \rangle^2} / {\langle 1/\pdet^2 \rangle} \gtrsim 0.1$, where the average is taken over the set of samples that were pre-selected and resulted in an exact S/N $>$ 8. In practice, Eq.~\eqref{eq:mean_SNR} seems to work quite well, and we can use very little training data (just 3000 samples of $\mathrm{S/N}(\theta, \phi)$ per detector sensitivity) and still achieve reweighting efficiencies $\eta_\mathrm{rw} > 0.8$. We have also explicitly compared this method against the brute force approach, and found no biases in the final set of detectable samples.

This way, we save on the number of wasted S/N computations that would result in values smaller than 8, resulting in computational savings of $\sim$ 1000x. Since the actual S/N are computed exactly, there are no unmodeled errors in the final result. If desired, this can be substituted with surrogate techniques~\citep{phurailatpam2024gwsnr, phurailatpam2024ler} for further speedup, though we often get sufficient samples with overall fewer S/N evaluations than would be needed to train these surrogates in the first place.

\section{Background and Foreground Simulations}
\label{sec:bgfg}

\begin{figure*}[t]
\centering
\includegraphics[width=\columnwidth]{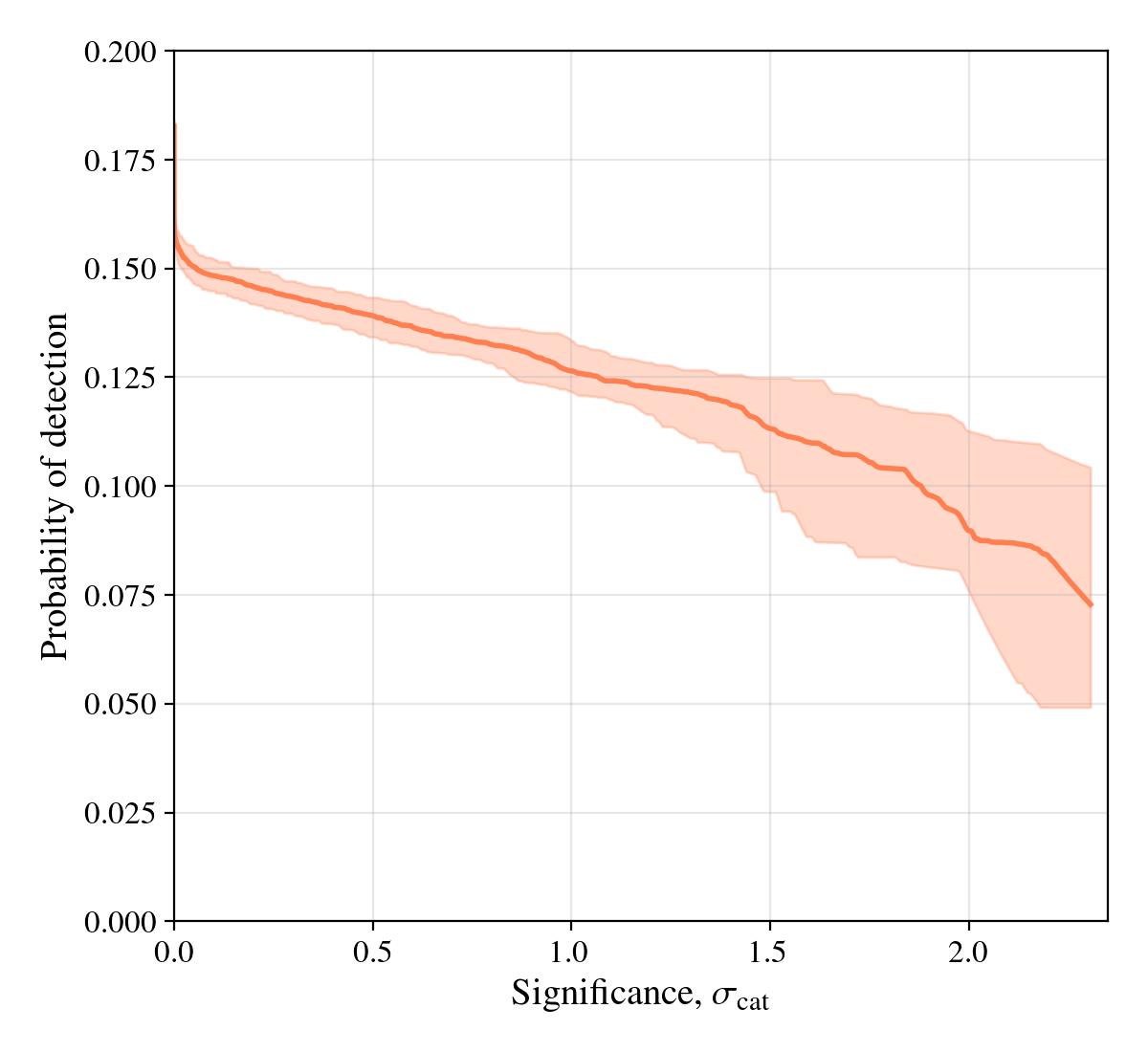}
\includegraphics[width=\columnwidth]{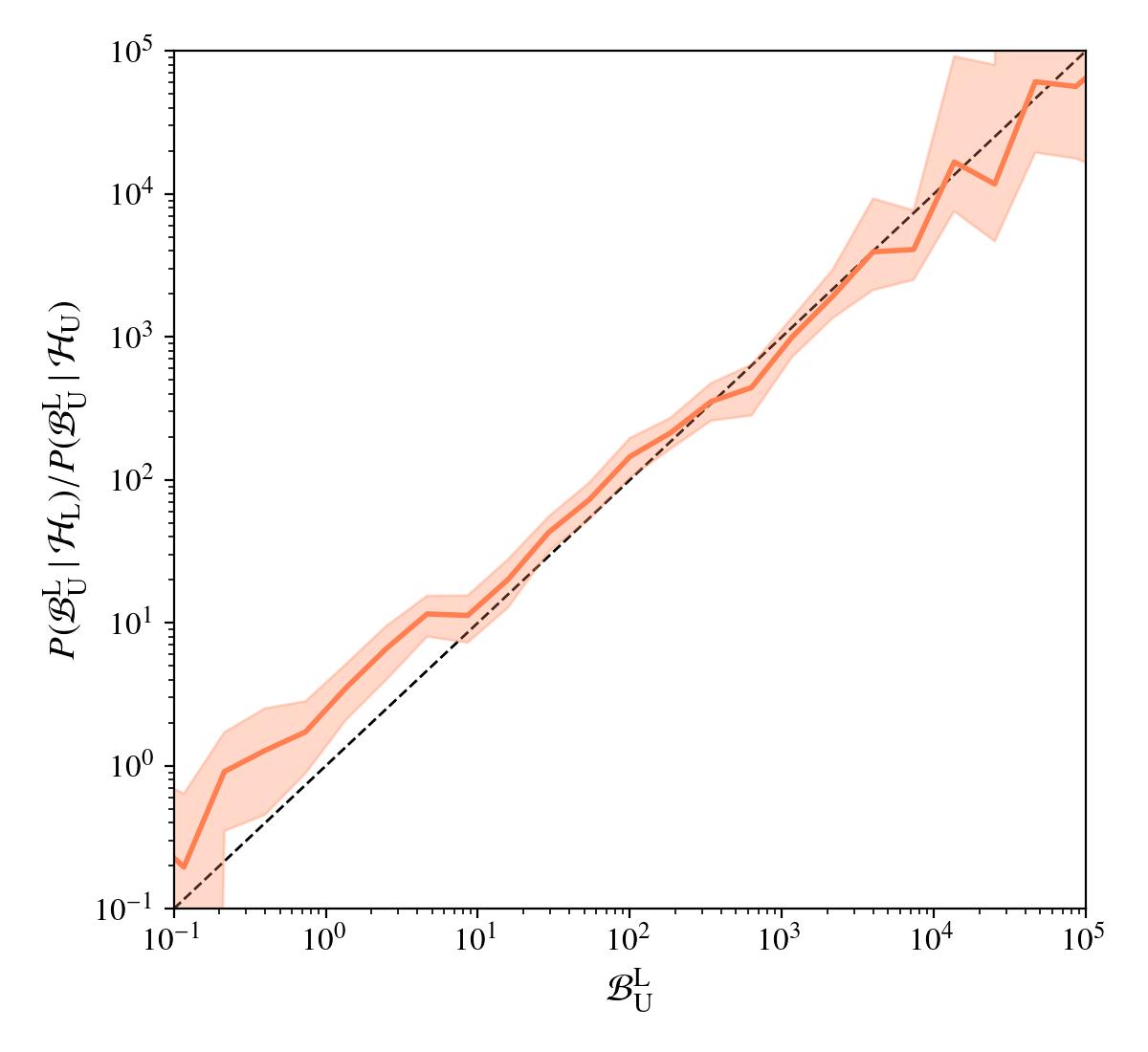}
\caption{\textit{Left}: The lensing-rate-included ROC curves for PO2.0 expressed in terms of the probability of making a lensing detection with a significance of at least $\sigma_\mathrm{cat}$. \textit{Right}: The \textit{B-B} plot for PO2.0, showing that the $\blu$ closely follows the ratio of its distributions under the lensed and unlensed hypotheses.}
\label{fig:performance}
\end{figure*}

Our background and foreground simulation consists of injections of detectable unlensed and lensed BBHs in real LIGO-Virgo detector noise from O1-O4a. We first identify time chunks during which CAT3 vetoed data was available, and further remove 32-second chunks around observed GWs. We then download randomly chosen, unique, 128-second-long real noise segments from \textsc{GWOSC}~\citep{abbott2021open, abbott2023open, abac2025open}, totaling about 160 hours for the Hanford and Livingston detectors, and about 107 hours for the Virgo detector. BBHs are injected in the downloaded segments that are closest to their arrival times. We inject in the same detectors that were operating at the simulated arrival time of the BBH, but since the segments themselves correspond to different time stamps, we make an assumption that the noise is incoherent across different detectors.

For the background, we draw a subset of 1500 BBHs from each of our generated unlensed priors, while for the foreground, we draw 1500 pairs of images of lensed BBHs from the prior. We then inject the corresponding \textsc{IMRPhenomXPHM} waveforms into real noise. For the saddle point images, we multiply the Fourier domain waveform by $i$. We clip the maximum in-band duration to 48 seconds, and leave a 32-second extra buffer on each side for PSD estimation and windowing.

During the prior generation, the detectability of the BBHs was assessed using the optimal S/N, which depends only on the noise PSD. For mimicking the real detection process more closely, we need to impose the threshold on the matched filter S/N, which fluctuates around the optimal S/N depending on the exact noise realization in which the signal is injected. We achieve this by randomly time-shifting the noise chunks until the matched filter S/N is found to be larger than 8.

We perform BBH parameter estimation on these 4500 injections using \textsc{Cogwheel}~\citep{roulet2022removing,islam2022factorized} with an \textsc{IntrinsicLVCPrior} within the frequency range of 20-1620 Hz. $\gtrsim 95\%$ of the PE runs were completed successfully, which we used for further analyses.

Apart from enabling us to define robust inference metrics, these background and foreground injections also help us validate and benchmark our pipeline. We plot the rate-included receiver operating characteristic (ROC) curves for PO2.0 in Fig.~\ref{fig:performance}. This is a plot of the probability of making a detection ($=1-e^{-uN ~ \fdp(\blu)}$) with a statistical significance of at least $\sigma_\mathrm{cat}(\blu)$ ($=\sqrt{2}~\mathtt{erfc}^{-1} ~ \fpp(\blu)$, where $\mathtt{erfc}^{-1}$ is the inverse of the complementary error function), plotted against $\sigma_\mathrm{cat}$. Our background of $\sim 1500$ allows us to probe $\sim 2.3\sigma_\mathrm{cat}$, at which the probability of making a detection is 5-14\%, at the combined sensitivity of O1-O4a runs, including the expected lensing rates.

We also make a \emph{Bayes factor-Bayes factor (B-B) plot}~\citep{barsode2026bb}, which is a general consistency check for Bayesian model selection pipelines. We find that, even in real detector noise, the PO2.0 Bayes factor is reasonably accurate at large values.

\section{Forecasting GW strong lensing detection}\label{sec:forecast}
We follow the method outlined in~\cite{barsode2026lensing} for forecasting the search efficiency at various $\fpp$ thresholds. Briefly, this involves simulating a population of detectable strongly lensed GWs in each observing scenario, converting their S/N and time delays to the strong lensing Bayes factor via a power law scaling relation, and using the \textit{B-B} relationship~(Eq.~20 of \cite{barsode2026bb}) to estimate the $\fpp$.

We assume a three-detector network consisting of LIGO Hanford, LIGO Livingston, and Virgo. For O4a, O4b, and IR1 scenarios we use the \textsc{O4high} noise curves taken from \cite{HLVK-psd-O3O4O5}. For the O5a, O5b, and O5c scenarios, we use \textsc{O5high} noise curve for Virgo, and the curves given in \cite{HL-psd-O5abc} for LIGO Hanford and LIGO Livingston. Their observing durations are set according to the current observing plans~\cite{igwn_obsplan_2025}.

Our results are estimated for three different astrophysical priors, differing in their assumptions of the high redshift merger rate and lens population. For one prior, we adopt a merger rate following the star formation rate~\citep{madau2014cosmic} with galaxy-scale lenses following the SDSS catalog~\cite{collett2015population}. This results in a low optical depth and is therefore pessimistic towards lensing. In the second prior, we also include the CBC formation efficiency~\citep{vitale2019measuring}. This shifts the peak source redshifts to higher values, resulting in larger lensing optical depths. Our third prior assumption is identical to that described in Sec.~\ref{sec:priors}. For each of these three priors, we allow the merger rate to vary between the current uncertainty of a factor of $\sim 1.6$. Finally, we summarize our forecasts in terms of the minimum, median, and maximum under these uncertainties in Fig.~\ref{fig:forecast}.

\bibliography{Bibliography}

\end{document}